%% file: CurveEvolution_arxiv_2016_resub.tex
\documentclass[aps,prx,reprint,superscriptaddress]{revtex4-1}
\usepackage{graphicx}
\usepackage{amssymb,amsmath}
\usepackage{epstopdf}
\usepackage{natbib}
\usepackage{array}
\usepackage{color}
\usepackage{xr}  

\newcommand{\eps}{\epsilon}

\include{commands_may11}

\begin{document}

\title{Designing steep, sharp patterns on uniformly ion-bombarded surfaces}
\author{Joy C. Perkinson}
\affiliation{Harvard School of Engineering and Applied Sciences, Cambridge, MA 02138}
\author{Michael J. Aziz}
\affiliation{Harvard School of Engineering and Applied Sciences, Cambridge, MA 02138}
\author{Michael P. Brenner}
\affiliation{Harvard School of Engineering and Applied Sciences, Cambridge, MA 02138}
\author{Miranda Holmes-Cerfon}
\affiliation{Courant Institute of Mathematical Sciences, New York University, New York, NY 10012}

\begin{abstract}
We propose and experimentally test a method to fabricate patterns of steep, sharp features on surfaces, by exploiting the nonlinear dynamics of uniformly ion-bombarded surfaces. 
We show via theory, simulation, and experiment, that the steepest parts of the surface evolve as one-dimensional curves which move in the normal direction at constant velocity. 
The curves are a special solution to the nonlinear equations that arises spontaneously whenever the initial patterning on the surface contains slopes larger than a critical value; mathematically they are traveling waves (shocks) that have the special property of being undercompressive. 
We derive the evolution equation for the curves by considering long-wavelength perturbations to the one-dimensional traveling wave, using the unusual boundary conditions required for an undercompressive shock, and we show this equation accurately describes the evolution of shapes on surfaces, both in simulations and in experiments. 
Because evolving a collection of one-dimensional curves is fast, this equation gives a computationally efficient and intuitive method for solving the inverse problem of finding the initial surface so the evolution leads to a desired target pattern.  We illustrate this method by solving for the initial surface that will produce a lattice of diamonds connected by steep, sharp ridges, and we experimentally demonstrate the evolution of the initial surface into the target pattern. 
\end{abstract}

\maketitle

\section{Introduction}

Fabricating steep, sharp features with desired morphologies on surfaces is a major challenge of materials science. 
Certain methods are available by direct engraving, such as focused ion beam (FIB) or lithography \cite{vasile97,adams03,li01,stein02}, but these require enormous time and energy. A promising method to efficiently make patterns on a large scale by exploiting dynamics is to erode a surface with uniform ion bombardment \cite{sigmund69,sigmund73,bradley88,chason,munoz2009}. A flat surface can become unstable and develop features such as quantum dots or hexagonal patterns \cite{facsko99,frost00,cuenat05,castro2005,wei2008,ziberi2008,munoz2010,bradley2010}. 
Such spontaneous pattern growth could spawn high-throughput methods to manufacture periodic metamaterials such as optical antenna arrays \cite{smythe2007} and the split ring resonators used in negative refractive index materials and optical cloaking \cite{rockstuhl2006}. 

However, because linear instabilities are neither small enough nor amenable enough to control, interest has recently turned to the potential for \emph{nonlinear} dynamics to create even steeper, sharper features \cite{chen2005}.  Large-amplitude, steep structures are of interest for three-dimensional engineering applications such as micromechanics, microprocessor integration, data storage, and photonic band gap waveguides.  They are also of interest in atom probe tomography, which uses samples frequently created by FIB ion irradiation \cite{miller2005} and which is strongly influenced by the geometry of the steep-walled final samples \cite{loi2013}.  Experiments and simulations have shown that knife-edge-like ridges, varying on scales at least an order of magnitude smaller than those accessible to linear instabilities, can arise spontaneously on \emph{uniformly} bombarded surfaces, provided the initial surface contains slopes beyond a critical value \cite{mhc12,mhc12b}. 
This suggests that one may be able to create steep, sharp features by first pre-patterning a surface on the macroscale -- something that is relatively easy to achieve -- and then bombarding it uniformly to let even steeper, sharper features form spontaneously.

In this manuscript we propose a theory to design patterns on a surface using their nonlinear dynamics, and we validate our approach experimentally. 
We argue that there is a dynamical regime where the full nonlinear dynamics can be approximated by evolving a collection of one-dimensional curves at constant speed in the normal direction, 
and we show this model accurately describes the experimental behavior of steep-walled pits propagating under uniform ion irradiation approximated via FIB rastering. 
This model can be used to solve the inverse problem of determining the initial surface pattern that will evolve under uniform ion bombardment to a desired target pattern, and as a demonstration, we numerically design and experimentally create a lattice where the scale of the lattice pattern is many times smaller than the scale of the initial patterning.

Our model has several advantages over a direct numerical simulation of the nonlinear equations for surface evolution.
First, it can rapidly determine how a given initial condition will evolve, e.g. by directly evolving the curves or using level set methods \cite{osher79,sethian96}, so the inverse problem can also be solved efficiently, for example using Monte-Carlo methods. 
Second, the full nonlinear equations require quantitative information about the macroscopic effects of uniform ion bombardment, such as the yield function (atoms ejected per incident ion as a function of angle of incidence) and the magnitude and type of smoothing physics, and this information is not always known nor easy to obtain \cite{madi2011,norris2011}. Our model requires knowing only one material-dependent parameter which can be measured through simple experiments, so measuring the full yield function and smoothing physics is unnecessary. 
Finally, this method is intuitive so an approximate guess for the initial pattern can be made without any sophisticated numerical techniques. 

\begin{figure*}
\begin{center}
\includegraphics[width=1\textwidth]{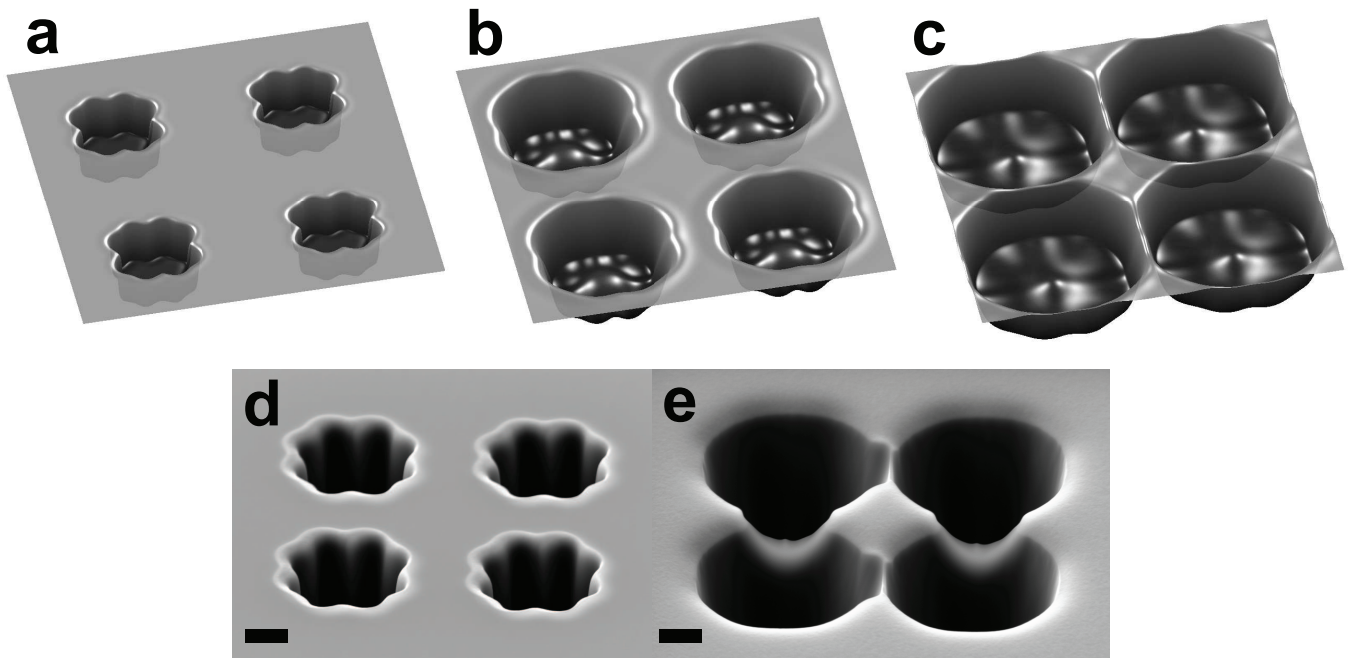}
\end{center}
\caption{ ({\bf a-c}) Numerical simulation of surface height evolution under uniform ion bombardment \eqref{eq:h}, at area doses 0, 3 nC/$\mu$m$^2$, and 6 nC/$\mu$m$^2$ (simulation times 0, 0.8, 1.6.)
 Pits initially spaced $O(1)\mu$m apart collide and form ridges with widths $O(100)$nm.  By changing the initial pattern one can create a large variety of patterns made of these steep, sharp ridges. ({\bf d}) SEM image of the fabricated initial condition predicted to evolve into knife-edge ridges, and ({\bf e}) the knife-edge ridges formed after ion irradiation in the FIB, both imaged at $54^\circ$ off-normal.  Scale bars are 1 $\mu$m.}\label{fig:sims}
\end{figure*}

\section{Results and Discussion}

\subsection{Evolving an Initial Condition to Produce a Target Morphology}

We begin by performing numerical simulations of the fully nonlinear, two-dimensional equations for the evolution of the surface height $h(x,y,t)$ under uniform ion bombardment, commonly described by the Sigmund theory of sputter erosion \cite{sigmund73}.   
When the surface slope varies over length scales much larger than the lateral scale over which an ion deposits its kinetic energy, the sputter integral can be expanded to yield the following nonlinear partial differential equation \cite{chen2005, aziz2006}: 
\begin{equation}\label{eq:h}
h_t + R(b)  + B_0\div\left(\oneover{\sqrt{1+b^2}} \grad \kappa\right) = 0.
\end{equation}
The first-order term $R(b)$ is the the average velocity of erosion of the surface as a function of its slope $b = |\grad h|$ (or equivalently the angle of the incoming ion beam). 
The fourth-order term with magnitude $B_0$ is a function of the surface curvature $\kappa = \div\left((1+b^2)^{-1/2} \grad h\right)$, which models additional smoothing effects such as Mullins-Herring surface diffusion \cite{mullins1959,herring1950} or ion-enhanced viscous flow confined to a thin surface layer \cite{umbach2001}.
Equation \eqref{eq:h} can also include second-order (curvature) terms \cite{bradley88, davidovitch2009}, but we neglect these as the nonlinear dynamics we wish to model can only occur when these are small \cite{mhc12b}.  

The erosion function $R(b)$ is related to the sputter yield $Y(b)$  by a constant of proportionality that changes its units from [atoms out/ions in] to [length/time].  This constant, as well as steep feature propagation speed, vary with ion flux.  Therefore, while we measure simulation progress in unitless simulation time, it is more informative to track experimental progress with area dose, a measure of ion fluence delivered per area reckoned in a plane parallel to the average surface.  When ion flux is held constant, area dose is proportional to simulation time.

We numerically simulate \eqref{eq:h} using the erosion function for 30 keV Ga$^+$ on Si (see \emph{Materials and Methods}), and estimate length scales for this material combination. 
Figure \ref{fig:sims}a-c shows simulations that begin with a periodic array of shapes punched out in the surface. 
The shapes are initially 2.5 $\mu$m apart (centers 5 $\mu$m apart) and are steep at their boundaries with nearly zero slope elsewhere. As the surface evolves, the steep regions remain steep while their location changes with time. 
Eventually the steep regions collide and form ridges that are even steeper, with widths of $\approx$ 100 nm. 
The resulting pattern is a lattice of diamonds connected by steep, narrow ridges.

These ridges were shown in \cite{mhc12b} to be a special solution to  \eqref{eq:h} that arises whenever steep regions propagating in opposite directions collide. The steepness and radius of curvature of this solution were shown to be fixed numbers that depend on the material, ion, and energy via $R(b)$, and we expect that  certain materials can achieve much smaller length scales \cite{mhc12}. Because identical ridges arise spontaneously, they are a useful structure to consider for patterns as they are not sensitive to the initial condition. 

To show that we can form a similar lattice of knife-edge ridges experimentally, we use FIB to mill four pits matching the initial shape used in simulations, then irradiate under repeated boustrophedonic FIB rastering delivering a low dose on each pass to approximate uniform irradiation, as described in {\it Materials and Methods}.  Due to discrepancies in pit wall propagation speed (see SI section \ref{sec:speed}), we milled the initial pits with centers 4.4 $\mu$m apart -- closer than in simulations.
Irradiation continued until the pits impinged (Figure \ref{fig:sims}d-e).
The resulting structure is close to that designed through our method and predicted by the simulations, 
 successfully demonstrating the formation of knife edge ridges under uniform ion irradiation. 
One difference from the simulations worth pointing out is that when pits evolved within 200 nm of each other, the pit rim closest to the adjacent pit accelerated and ``reached out'' to the adjacent pits.  These behaviors caused impingement to occur both sooner than expected and over a smaller length of the rim.  Bombardment was continued after this impingement in order to extend the lateral dimensions of the knife edges, and so the knife edge evolved into a curved ridge. Indeed, our simulations show that if bombardment continues beyond the moment of impingement, the knife-edge ridges may evolve into the curved shaped similar to those observed in experiment (SI Figure \ref{fig:curvedfig}).

\subsection{One-dimensional curve evolution equations}

We designed the pattern of ridges by solving an inverse problem for the initial conditions, and we now explain these calculations. The basic observation is that
 we can draw a curve through the steepest parts of the surface at every time step, and watch the curve  evolve in time. Our goal is to derive an evolution equation for the curve using only the curve's intrinsic, one-dimensional geometry, and not any information away from the region of high gradient.
  
 To look for such an equation we recall the theory developed in \cite{mhc12}. This showed that Eqn. \eqref{eq:h} has a particular traveling wave solution for the slope that is invariant in one horizontal dimension, i.e. of the form $h_x = s(x-ct)$,
that acts as an attractor for the dynamics: if a surface is patterned initially to have slopes above a critical value, then the surface steepens and locally evolves to that traveling wave. This wave is called \emph{undercompressive} in the mathematical literature \cite{bertozzi99,bertozzi2001,lefloch2008} because it has the non-classical property that information can propagate away from it. 
In our system it is identifiable because it connects a steep region with constant slope $b_0 = s(-\infty)$ to a flat region with slope $0=s(\infty)$, and it propagates at constant speed  $c = (R(b_0)-R(0))/(b_0-0)$, where the constants $b_0, c$ depend on the material, ion, and energy via $R(b)$ but not on the initial condition. When we extract the slope and speed in the simulations of shapes such as in Figure \ref{fig:sims}, we find that they are close to $b_0, c$, suggesting that the surface slope can be locally approximated as a collection of undercompressive traveling waves with slowly varying phase shifts in the transverse direction.

To understand how this collection evolves we consider two different theories. 
\emph{Theory 1}  simply advects curves in the normal direction with speed $c$. If the curve is parameterized by $\alpha$ as $q(\alpha,t)=(x(\alpha,t),y(\alpha,t))$,  it evolves according to 
\begin{equation}\label{eq:phase1}
\dd{q}{t} = c\hat{n}, 
\end{equation}
where $\hat{n} = \pp{q}{\alpha}^\perp / |\pp{q}{\alpha}|$ is the unit normal to the curve. This theory requires only one parameter, the wave speed $c$, which can be measured by simple experiments. For example, one can engrave a circular pit and measure the rate of change of its radius, as in \cite{mhc12b}. 
Theory 1 is a natural heuristic, and is expected to be valid when 
the steep parts of the surface form a curve that is nearly straight, so it can be treated locally as a one-dimensional traveling wave.

\emph{Theory 2} seeks to describe the evolution of transverse perturbations to a one-dimensional traveling wave, by looking for an asymptotically consistent solution to the nonlinear, two-dimensional equations. 
We start by 
looking for a solution whose slope has the form
\begin{equation}\label{eq:ansatz}
h_x = s(x-ct + \psi(y, t)) + \epsilon u(x-ct,y,t) + O(\epsilon^2).
\end{equation}
We assume the scalings $\partial_y \sim O(\eps^{1/2})$, $\partial_t \sim O(\eps)$, with $\epsilon\ll1$. 
We substitute this ansatz into the $x$-derivative of \eqref{eq:h} and perform a multiscale asymptotic expansion \cite{BenderOrszag}.
At leading order is the equation for $s$, which is satisfied by construction. 
The $O(\epsilon)$ equation is 
\begin{multline}\label{eq:u}
u_t + \mathcal{L}u =
 a_1(s)\psi_{t} + a_2(s)\psi_{y}^2 + a_3(s)\psi_{yy} + a_4(s)\psi_{yyyy} 
\end{multline}
where $a_i(s)$ are functions of the traveling wave (SI, eqns. \eqref{eq:a1}-\eqref{eq:a4}) and the linear operator $\mathcal{L}$ is 
\begin{multline}\label{eq:L}
\mathcal{L}u =  \partial_\eta((R'-c)u) + \\
B_0\left[\partial^2\eta(f_d\partial^2_\eta(sf)u)
+ \partial^2_\eta(f\partial^2_\eta((\half f_ds^2+fs)u))\right].
\end{multline}
We write $\eta=x-ct$, $f(b) = (1+b^2)^{-1/2}$,  $f_d(b) = \dd{}{b}f(b)$, and all functions are evaluated at $s(\eta)$. 
We have included the term proportional to $\psi_{yyyy}$ from the $O(\epsilon^2)$ equation because it is sometimes required to smooth (SI, section \ref{sec:terms}.) 

The left-hand side of \eqref{eq:u} depends only on the fast variables $\eta,t$, so we can integrate over these to derive a solvability condition. Suppose there is a function $\pi(\eta)$ such that $\mathcal{L}^*\pi = 0$, where $\mathcal{L}^*$ is the adjoint of $\mathcal{L}$ with respect to the $L_2$-inner product $\langle \cdot, \cdot \rangle$ (i.e. it satisfies $\int_x u(\mathcal{L}v) dx = \int_x (\mathcal{L}^*u)vdx$ for all $u,v$ with the appropriate boundary conditions.) Taking the inner product  with \eqref{eq:u} and requiring $u$ to be bounded 
shows that $\langle \pi, RHS\rangle = 0$, where RHS is the right-hand side of the equation. Therefore the phase will evolve on the slow timescale as
\begin{equation}\label{eq:phase2}
\psi_{t} + c_2\psi_{y}^2 + c_3\psi_{yy} + c_4 \psi_{yyyy} = 0,
\end{equation}
where $c_i = \langle \pi, a_i(s) \rangle / \langle \pi, a_1(s) \rangle$. 

To find $\pi$ requires solving $\mathcal{L}^*\pi = 0$ with the appropriate boundary conditions, which are  $\pi(-\infty) = 0$, $\pi(\infty) = 1$ for the undercompressive traveling wave \cite{bertozzi2001,liuzumbrun95,sattinger76,zumbrun1999}. Using these one can compute $\pi$ numerically, and then the constants can be found by numerical integration. 
The condition of decay at $-\infty$ is unusual and is what makes this multiscale analysis novel. The condition arises in order to control information that can propagate away from the traveling wave on its undercompressive side (SI, section \ref{sec:bdy}.) 

Equation \eqref{eq:phase2} forms the basis of Theory 2. It demonstrates rigorously that the nonlinear dynamics of an ion-bombarded surface can be approximated (for long-wavelength pertubations) as the evolution of a collection of curves on the surface, each one propagating at constant speed $c$ in a certain direction and changing shape about this direction according to \eqref{eq:phase2}. The theory requires four parameters: $c_2,c_3,c_4$, and $c$. These can be calculated numerically if the erosion rate $R(b)$ and the magnitude of the fourth-order term $B_0$ are known for a given material. 
If they are not known, they could be extracted from experiments which measure the evolution of different shapes. 

Note that Theory 2 considers perturbations about a horizontal reference line, so it relies on a particular coordinate system. Theory 1 on the other hand is intrinsic: it depends only on the local geometry of the traveling wave front. We expect that one could derive Theory 1 from the governing equations \eqref{eq:h} by considering a slowly varying traveling wave, and additionally that one could derive the next-order corrections to \eqref{eq:phase1} as we have done in Theory 2. We do not do this here because, as we will show, the difference between the predictions of the two theories is so small as to be undetectable experimentally, but this would be an interesting question for future analysis.

\begin{figure}
\center
\includegraphics[width=\linewidth]{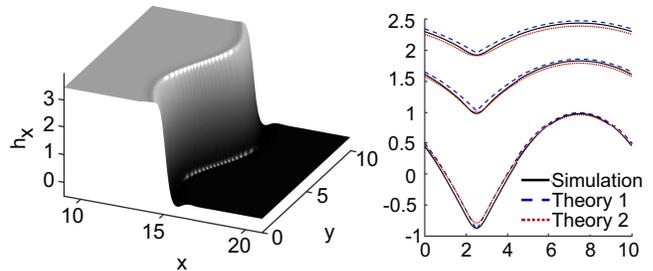}
\caption{
Numerical tests comparing the curve evolution equations \eqref{eq:phase1},\eqref{eq:phase2} to simulations of the full dynamics \eqref{eq:h}. 
Left: initial surface slope $h_x(x,y,0) = s(x + \sin(\frac{2\pi}{10} y))$. 
Right: curve extracted from full simulation (solid line), curve predicted by \eqref{eq:phase1} (blue dashed) and curve predicted by \eqref{eq:phase2} (red dotted), at times 2,7.5,12 (corresponding to area doses 7.5, 28.125, and 45 nC/$\mu$m$^2$).
The curves at each different time have been plotted at 1/10 the actual separation in the vertical direction. 
}\label{fig:phase}
\end{figure}

\subsection{Comparing theories, simulations, and experiments}

To test how well Eqns. \eqref{eq:phase1},\eqref{eq:phase2} describe the propagation of steep features, we compared them with numerical simulations of the full two-dimensional equations \eqref{eq:h}. 
We started with a traveling wave computed as a steady solution to the discretized version of \eqref{eq:h}, applied a sinusoidal perturbation to the surface slope in the transverse direction, evolved this surface numerically, and identified the curve by the maximum of $|h_{xx}|$ as a function of $y$ at each timestep. We compared this curve to numerical simulations of \eqref{eq:phase1},\eqref{eq:phase2} with the same sinusoidal initial condition (\emph{Materials and Methods}.) 

Figure 2 shows the three curves at different times. The curves predicted by the theories agree extremely well with the curve extracted from the simulations. This agreement is destroyed when the parameters are changed from their predicted values, so it is not an accident. 
The small discrepancies between theoretical predictions and simulations are thought to come from two sources: numerical discretization of the two-dimensional equations, and higher-order asymptotic corrections to the theoretical curves. 
It is notable that the curves predicted by both theories are also extremely close to each other, showing that while they make different kinds of approximations, they may be used roughly interchangeably. Therefore, Theory 1 should be preferred under the conditions investigated here, because it is simpler.

\begin{figure}
\includegraphics[width=\linewidth]{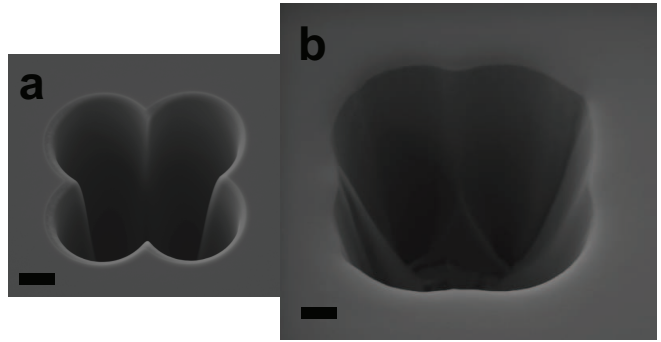}
\caption{
({\bf a}) Initial milled ``clover'' pit and ({\bf b}) milled pit after an area dose 30 nC/$\mu$m$^2$
imaged under SEM at $30^\circ$ off-normal.  Scale bars are 2 $\mu$m and identical.
}\label{fig:cloverexp}
\end{figure}

\begin{figure}
\includegraphics[width=\linewidth]{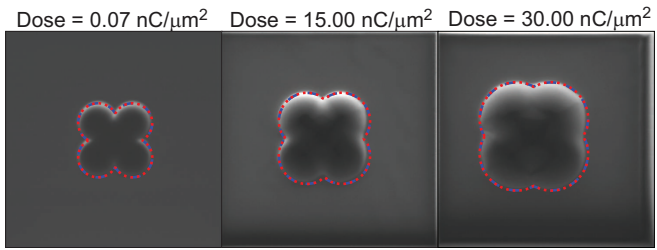}
\caption{
The evolution of the clover-shaped pit under homogeneous FIB rastering of the entire imaged area, for irradiation doses indicated. Theoretical predictions using equations \eqref{eq:phase1} (blue) and \eqref{eq:phase2} (red) are superimposed. 
}\label{fig:cloversims}
\end{figure}

We then compared experimental pit propagation to the predictions of Theories 1 and 2. 
We started with a clover-shaped hole, formed by milling 
four overlapping circular pits (Figure 3)  with radius 2.9 $\mu$m, centered at ($\pm2.4$ $\mu$m, $\pm2.4$ $\mu$m).  A fifth pit with radius 1.5 $\mu$m was milled at the origin to remove the extra material not removed by the other four pits.  All pits were milled in parallel in order to minimize the effects of Si redistribution on pit walls.  The initial and final pit morphologies, imaged using scanning electron microscopy (SEM) at $30^\circ$ off-normal, are shown in Figure 3.  The horizontal pit width expands from 10.8 $\mu$m to 15.7 $\mu$m after an area dose of 30 nC/$\mu$m$^2$ Ga$^+$.

We simulated the evolution of a curve using Eqns. \eqref{eq:phase1},\eqref{eq:phase2} with initial condition set to the boundary of the clover-shaped hole. 
We fit the propagation speed $c$ to that observed experimentally, since we were unable to match it quantitatively from first principles (SI section \ref{sec:speed}.) 
The simulated curves are overlaid on the real-time FIB images for comparison and shown after an area dose of 0.07 nC/$\mu$m$^2$, 15.00 nC/$\mu$m$^2$, and 30.00 nC/$\mu$m$^2$ in Figure 4.
Both simulations agree equally well with experiment, overlapping at the pit walls with greatest radius of curvature.  At the four ``kinks'' with tighter radius of curvature, the simulated curves vary from experiment by an average of 0.1 $\mu$m.  Both models could be used essentially indistinguishably in these experimental systems.

\subsection{Solving the inverse problem}

Our numerical simulations and experimental tests show that either Eqn. \eqref{eq:phase1} or Eqn. \eqref{eq:phase2} can be used to predict the propagation of steep regions on the surface.  
These are intuitive equations that make it easy to sketch by hand an approximate initial condition for a surface that evolves under bombardment to a given final pattern.  
In addition, because evolving a collection of curves is fast, the inverse problem can be efficiently solved more precisely by numerical methods, for example by Monte Carlo simulations. To illustrate, we explain how we designed the lattice in Figure \ref{fig:sims}a.  Our target pattern is a periodic array of tiles shown in Figure 5a. The black pixels are regions where we wish the surface to be elevated, where the borders are intended to be the knife-edge ridges. We will make this pattern using a periodic array of curves, each of the polar form  $r(\theta) = r_0(1 + Re\left\{ \sum_{k=-k_{max}}^{k_{max}} w_ke^{ik\theta} \right\})$, where $\{w_k\}_{k=-k_{max}}^{k_{max}}$ are parameters to be determined. We set $k_{max}=8$ and require the pattern to be formed at a fixed simulation time $T$. 
These restrictions do not come from realistic experimental constraints, but rather are intended to illustrate the more general principle that one can optimize over a constrained set of initial conditions. 

We then used Monte Carlo simulations to find the initial conditions that lead to the desired pattern.
At each Monte Carlo step we varied one parameter, solved \eqref{eq:phase2} up to time $T$, and computed the cost as the sum of the absolute discrepancy between the set of pixels lying outside each closed curve, and the target pattern. We discarded moves that increased the cost with a cost-dependent probability. 
Figure 5b shows the optimal initial condition after a large number of Monte Carlo steps (blue) and the final curve that it evolves to (red).  Figure 1a-c shows the evolution of this initial condition with a simulation of \eqref{eq:h}; again there is excellent agreement.

\begin{figure}
\includegraphics[width=\linewidth]{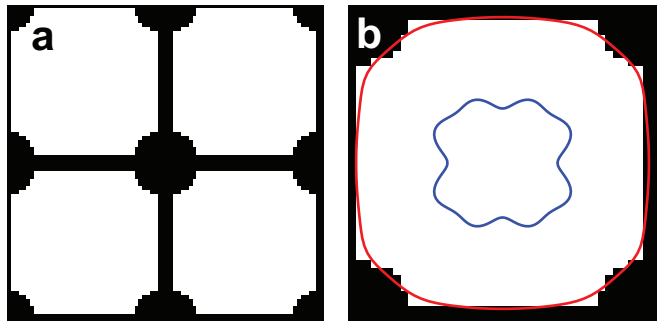}
\caption{
({\bf a}) Target pattern used to design the lattice in Figure \ref{fig:sims}, where black pixels are elevated regions. ({\bf b}) Initial condition with lowest cost (blue) and curve it evolves to (red). The cost of this initial condition is the sum of the number of white pixels lying outside the red curve, plus the number of black pixels lying inside it.
}\label{fig:mc}
\end{figure}

\section{Conclusion}
We have introduced a method to make steep, sharp patterns on surfaces, by pre-patterning the surface  so it dynamically evolves under uniform bombardment to something that is much smaller-scale and more difficult to make directly. 
Our method is based on the demonstration that for certain materials, ions, and energies, the 
steep parts of the surface evolve as one-dimensional curves which propagate at a constant speed in the normal direction. 
This simplification agrees extremely well with simulations of the full nonlinear dynamics, 
and over large enough scales it also agrees with experimental measurements of the evolution of steep-walled pits under rastered FIB irradiation. On small scales, such as when steep features approach each other to form sharp ridges, the theory is not expected to apply because the governing equations are based on a small-curvature approximation, and indeed on these scales we observed several phenomena that we do not yet understand. 
In addition, we were unable to quantitatively predict the speed of the curve from first-principles, but this is easy to measure and then incorporate into the theory. 
 Therefore, we propose that this method of evolving one-dimensional curves may be used as a first approximation to determining a surface's final structure after uniform bombardment. 
 
Evolving curves is fast, so it naturally leads to efficient methods to solve the inverse problem of determining how to pattern the surface initially so it evolves to a desired target structure. 
We showed how to make a simple lattice connecting diamonds with steep ridges, but one can imagine making more complex patterns, for example by starting with several different shapes, non-closed curves, bumps instead of pits, or undulating topography, and also by allowing the curves to continue evolving once they intersect, to form gaps.  The space of possibilities is large and we expect that further understanding of the nonlinear dynamics of ion-bombarded surfaces will lead to new methods to invent and fabricate materials.

\medskip
\noindent

\section{Materials and Methods}
\subsection{Experimental Methods}
Samples were polished (001) Si wafers from Virginia Semiconductor, Inc. and were irradiated using a ZEISS NVision 40 focused ion beam (FIB) using NanoPatterning and Visualization Engine (NPVE) software.  Samples were affixed to aluminum sample stubs using silver paste, and dust was removed using an air jet prior to loading into the FIB.  The scanning electron microscope (SEM) beam was switched off during FIB irradiation to avoid the surface carbon contamination typical of SEM imaging.  The ion beam was 30 keV Ga$^+$, with a current of 1.5 nA, beam diameter 200 nm, center-to-center dot spacing 100 nm, and dwell time 1.0 $\mu$s.  Steep-walled pits were milled by rastering the beam repeatedly, each time delivering a small dose to the pit area and sputtering away a thin layer of Si.  The beam was rastered over the milling area at least 10000 times to create each pit, ensuring that the effects of non-uniform Si redistribution were minimized.   Circular pits were created by rastering the beam in circles from the center outward to the outer rim to create a maximally clean, steep pit wall.  Circular pits were overlapped to form the ``clover'' shapes.
	To cause pit evolution, a square bombardment area was chosen to overlap the pits.  The same 1.5 nA, 30 keV FIB beam was then repeatedly rastered over this area in a double serpentine (boustrophedonic followed by its time-reverse) scan, delivering a small dose in each pass to approximate uniform irradiation.  A depth of approximately 0.5 nm of material was sputtered away with each pass of the beam.  Images of the pit shape were captured during irradiation using FIB imaging using the secondary electron (SE2) detector.  
Initial and final morphology were imaged using SEM.

We found that the pit wall propagation rate was slower than the theoretically-predicted rate; to counteract this discrepancy and ensure impingement occurred at the correct moment during pit evolution, the initial pits were milled closer than directed by the simulations.

\subsection{Numerical Methods}

All numerical simulations were performed using an erosion function for 30 keV Ga$^+$ on Si, found in \cite{chen2005} to be
\begin{multline}\label{eq:R}
R(b) = R(0)\frac{\sqrt{1+b^2}}{\sqrt{1+b^2\mu^2/\sigma^2}} \times \\
\text{exp}\left\{ -\frac{a^2/\sigma^2}{2(1+b^2\mu^2/\sigma^2}  + \frac{a^2/\sigma^2}{2} - \Sigma(\sqrt{1+b^2}-1)\right\}
\end{multline}
with parameters $a/\sigma = 2.04$, $\mu/\sigma=0.658$, $\Sigma = 0.0462$.

To simulate \eqref{eq:h}, we non-dimensionalized the equation by scaling lengths by $L = (B_0/R_0)^{1/3}$, times by $T=L/R_0$, and the yield function by $R(b)\to R(b)/R_0$ where $R_0 = R(0)$. We then used a non-dimensional parameter $\tilde{B}_0 = 0.02$ so we could simulate scales much large than the width of the traveling wave. We used a semi-implicit numerical method introduced in \cite{mhc12b}. 
The discretization changes the undercompressive slope and speed from their theoretical non-dimensional values of $b_0=3.89$, $c=1.73$  to $b_0=3.45$, $c=1.84$. We used the numerical values in \eqref{eq:phase1}, \eqref{eq:phase2} when comparing to simulations. 

We simulated \eqref{eq:phase2} using the same semi-implicit method as \cite{mhc12b} but applied to a one-dimensional equation. For radial pits we assumed the phase $\psi$ was a perturbation to the radius $r(t)$ so changed to radial coordinates by making the substitution $\partial_y \to r^{-1}\partial_y$. Coefficients $c_i$ were calculated numerically by first computing the undercompressive traveling wave solution $s(\eta)$ using Matlab's bvp solver, then computing $a_i(s)$ using centered differences for the derivatives, and finally computing $\pi(\eta)$ as the second element in the null space of the numerically discretized version of $\mathcal{L}$ in \eqref{eq:L} (the first element of the discretized operator is always constant.) We performed an affine transformation on $\pi$ to ensure it had the correct boundary conditions since these are not required for the discretized operator. 
The traveling wave and coefficients were calculated for the nondimensional equation and were re-dimensionalized using $B_0 = 0.02$ to compare to simulations and experiments. 
The non-dimensional values were $c_2 = 0.866$, 
$c_3 = -0.245$, 
$c_4 = 0.231$. 
To re-dimensionalize we multiply $c_2$ by $L/T = R_0$, $c_3$ by $L^2/T = R_0^{2/3}B_0^{1/3}$, and $c_4$ by $L^4/T = B_0$ (Note that $\psi$ has units of length.) 
For the experiments, we do not know the true value of $B_0$, but as long as it is small it makes little difference to the curve dynamics. 

We simulated \eqref{eq:phase1} by discretizing the curve, calculating tangent vectors with centered differences, and updating each point on the curve according to \eqref{eq:phase1}. 
To prevent the curve from self-intersecting we added a small term  $\eps \kappa_1$ to the right-hand side, where $\kappa_1$ is the curvature vector, calculated using centered differences on the normalized tangent vectors. When the minimum separation between the points parameterizing the curve dropped below a threshhold we re-parameterized, by linear interpolation. This step provides a smoothing that in some cases was sufficient to prevent the curve from self-intersecting, so we could use $\eps=0$. Otherwise, we chose 
$\eps=1\times10^{-5}$.  This was small enough that the evolution was indistinguishable by eye from a curve which evolves with $\eps=0$ over the regions that have not yet self-intersected. 

\bigskip

\begin{acknowledgments}
This research was funded by the National
Science Foundation through  the Harvard Materials Research Science and Engineering
Center DMR-1420570 and  the Division of Mathematical Sciences DMS-1411694.
MPB is an investigator of the Simons Foundation. J.C.P. and M.J.A. were supported by NSF-DMR-1409700.  This work was performed in part at the Center for Nanoscale Systems (CNS), a member of the National Nanotechnology Infrastructure Network (NNIN), which is supported by the National Science Foundation under NSF award no. ECS-0335765. CNS is part of Harvard University.
\end{acknowledgments}

\bibliography{ShockBib.bib}


\appendix

\section{Speed calculations}\label{sec:speed}

According to the theory, the speed $c$ of the curves is uniquely determined by the yield curve and the type of smoothing physics (but not its magnitude). 
We compared theory to experiments, starting from the previously measured yield curve and our best knowledge of the smoothing physics, believed to be surface diffusion. 

In Figure 4, shocks were measured moving 80 nm over a delivered area dose of 1 nC/$\mu$m$^2$.  For the flux used in that experiment, this area dose corresponds to a time of 602 s.  Thus, the experimental shock propagation speed is found to be 80 nm / 602 s = 0.13 nm/s.

The shock speed is theoretically given as $c=(R(b_0)-R(0))/(b_0-0)$, and depends on the fixed slope $b_0$ and the erosion velocities $R(b_0)$ and $R(0)$.  The slope is $b_0=3.89$ as described by Chen et. al. \cite{chen2005}.  Theoretical values for the erosion velocities $R(b)$ can be derived using the sputter yield $Y(b)$, a measure of atoms sputtered away from the surface per incident ion, by changing dimensions.  
The sputter yield for normal incidence ions, $Y(0)=2.78$, is found using SRIM simulations and reported by Giannuzzi {\it et al.} \cite{giannuzzi}, which along with the equation for the angular dependence of the normalized sputter yield modified by the empirical Yamamura correction factor \cite{yamamura1983, yamamura1987} yields a value of $Y(b_0)=21.5$.  The dimensional erosion velocities can be calculated using these sputter yields and the atomic volume of silicon, $2.00\times 10^{-29}$ m$^3$:
\begin{widetext}
$R(0) = \frac{1\textrm{ nC}/\mu \textrm{m}^2}{602\textrm{ s}}\left( \frac{6.24\times 10^9\textrm{ ions in}}{1\textrm{ nC}}\right) \left( \frac{2.78\textrm{ atoms out}}{\textrm{ions in}}\right) \left( \frac{2.00\times10^{-29}\textrm{ m}^3}{\textrm{atoms out}}\right) \left( \frac{10^{18}\textrm{ }\mu\textrm{m}^3}{1\textrm{ m}^3}\right)\left( \frac{10^3\textrm{ nm}}{1\textrm{ }\mu\textrm{m}}\right) = 0.576 \textrm{nm/s}
$

$R(b_0=3.89) = \frac{1\textrm{ nC}/\mu \textrm{m}^2}{602\textrm{ s}}\left( \frac{6.24\times 10^9\textrm{ ions in}}{1\textrm{ nC}}\right) \left( \frac{21.5\textrm{ atoms out}}{\textrm{ions in}}\right) \left( \frac{2.00\times10^{-29}\textrm{ m}^3}{\textrm{atoms out}}\right) \left( \frac{10^{18}\textrm{ }\mu\textrm{m}^3}{1\textrm{ m}^3}\right)\left( \frac{10^3\textrm{ nm}}{1\textrm{ }\mu\textrm{m}}\right) = 4.456 \textrm{nm/s}
$
\end{widetext}

The shock speed from theory is thus $c=(4.456\textrm{ nm/s}-0.576\textrm{ nm/s})/(3.89-0)=0.997$ nm/s, a factor of 8 larger than that measured from experiment.  

This discrepancy means that either there is something wrong with the yield curve, or there is smoothing physics that is not yet incorporated in the model. We do not understand this discrepancy and leave it a question for future research. Note however that this is not important for the theory in this paper, since for the theory to work what matters is that the speed and slope are uniquely selected. We can simply measure the shock front velocity by evolving circular pits, and use this measured velocity to evolve shapes.





\bigskip


\subsection{Knife edge ridge curving after initial formation}\label{sec:curvedwall}

\begin{figure*}
\center
\includegraphics[width=0.8\textwidth]{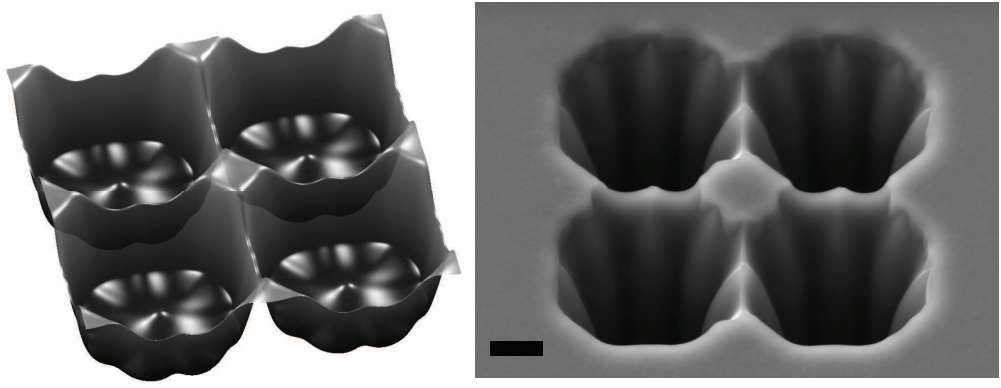}
\caption{Left: simulated knife edge ridge behavior if irradiation continues after steep feature impingement.  The initial hole had a non-dimensional depth of -4, and the simulation was run until time $1.8$. Right: SEM image of curved pit walls resulting from irradiation after steep feature impingement, viewed at 30$^\circ$ off-normal.  Initial pits were milled with centers 3.4 $\mu$m apart, and irradiated with an area dose of 1.2 nC/$\mu$m$^2$.}\label{fig:curvedfig}
\end{figure*}

Simulation and experiment both demonstrate that knife edge ridges become curved if irradiation continues after the initial impingement of steep features, as shown in Figure \ref{fig:curvedfig}.  The shape of the wall, with a central high point, arises due to the slight curvature remaining in the shape of the propagating walls at the moment of impingement, which causes the midpoint of the ridge to form later than the rest of the ridge.  This central high point is absent from the ridges shown in Figure 1e of the manuscript because initial pits were spaced further apart, resulting in a more uniformly-round shape at the moment of bombardment.  Furthermore, because the pits ``reached out'' to each other when they evolved within 200 nm of each other, the center of the ridge formed first, thus removing the source of the central high point.

\begin{widetext}

\section{Detailed multiscale calculations}\label{sec:epsilons}

In this section we record the details of the multiscale expansion used to derive equation [{\bf 6}],
the basis of Theory 2. 
We work in the frame of reference of the traveling wave by defining $\eta = x-ct$. The $\eta$-derivative of [{\bf 1}] 
is
\begin{equation}
(h_\eta)_t + (R(b))_\eta  - ch_{\eta\eta}+ B_0 \partial_\eta\div(f(b)\grad\div(f(b)\grad h))
\end{equation}
with $f(b) = (1+b^2)^{-1/2}$, $b=|\grad h| = \sqrt{h_x^2+h_y^2}$. We make the ansatz
$h = S(x-ct+\Psi(y,t)) +h_0(t) + \epsilon h_1 + \epsilon^2 h_2  + \ldots$ where $s(\eta) = S'(\eta)$ is the traveling wave solution, and assume the scalings $\partial_t \sim O(\eps)$, $\partial_y \sim O(\eps^{1/2}).$ 

The $O(1)$ and $O(\eps)$ parts of various terms are shown in the table below. 
The fourth-order term has been broken up by first calculating  $k(b) = \div(f(b)\grad h) = \partial_\eta(fh_\eta) + \partial_y(fh_y)$, and then calculating $m(b) = \partial_\eta\div(f(b)\grad k(b))$. 

To distinguish between various derivatives, we write a subscript ``d'' when we mean the pointwise derivative of a function,with no chain rule involved, i.e. $f_d(s(\eta)) = \dd{}{s}f|_{s(\eta)}$. We write a $'$ to mean derivative with respect to $\eta$, i.e. $f'(s) = f_d(s)s'(\eta)$. All functions are evaluated at $s$. 

\bigskip

\begin{tabular}{>{$}p{1.5cm}<{$}>{$}p{2cm}<{$}>{$}p{1.6cm}<{$}>{$}p{8cm}<{$} }
 & O(1) & O(\eps^{1/2}) & O(\epsilon)  \\
h&  S + h_0(t) && h_1  \\
h_t & s + \dd{}{t}h_0 &&  s\psi_t + h_{1,t} \\
h_{\eta} & s && h_{1,\eta}\\
h_{\eta\eta} & s' && h_{1,\eta\eta}\\
h_y && s\psi_y \\
h_{yy} && &\psi_y^2s' + s\psi_{yy} \\
b & s && \half s\psi_y^2 + h_{1,\eta}  \\
f(b) & f(s) && f_d(\half s\psi_y^2 + h_{1,\eta}) \\
f_y && f'\psi_y \\
(R(b))_\eta & (R(s))' && (\half R_d s)'\psi_y^2 + (R_d(s))'h_{1,\eta} + R_dh_{1,\eta\eta}\\
k(b) & (fs)' && \partial_\eta[(f_ds+f)h_{1,\eta}] + fs\psi_{yy} + (\half f_ds^2+fs)'\psi_y^2 \\
\kappa_y && (fs)''\psi_y\\
\\
m(b) & (f(fs)'')'' && = \partial_{\eta\eta}(f_0k_{1,\eta} + f_1k_{0,\eta}) + \partial_\eta\partial_y(f_0k_{0,y}) \\
  &&&\\
  & &&=\psi_{yy}\cdot  \left( (f(fs)'')' + (fs)''' \right) \\
  &&& + \; \psi_y^2 \cdot\left(  (\half f_ds(fs)'')'' + (f(fs)'')'' + (f(fs')')'' + (f(\half f_ds^2)'')''   \right)\\
  &&& + \; (f((f_ds+f)h_{1,\eta})'')'' + (f_d(fs)''h_{1,\eta})''\\
\end{tabular}\\

Some auxiliary calculations:\\

\begin{tabular}{>{$}p{1.7cm}<{$} c >{$}p{10cm}<{$} }
f_0k_{1,\eta} &= & f\partial_{\eta\eta}[(f_ds+f)h_{1,\eta}] + f(fs)'\psi_{yy} + f(\half f_ds^2+fs)''\psi_y^2\\
f_1k_{0,\eta} &= & \half f_d s(fs)''\psi_y^2 + f_d(fs)''h_{1,\eta} \\
\partial_y(f_0k_{0,y}) &=& (f(fs)'')'\psi_y^2 + f(fs)''\psi_{yy}\\
\end{tabular}\\
\bigskip

Collecting up terms gives the $O(1)$ equation
\begin{equation}
\partial_{\eta\eta}(f(s)\partial_{\eta\eta}(f(s)s)) + (R(s))' - cs'  = 0.
\end{equation}
This is satisfied by construction, since $s(\eta)$ is assumed to be a solution. 

Next we collect up terms for the $O(\epsilon)$ equation, and also include the term for $\psi_{yyyy}$ from the $O(\eps^2)$ equation. This gives, with $u=h_{1,\eta}$:
\begin{equation}
u_t + \mathcal{L}u = a_1(s)\psi_t + a_2(s)\psi_y^2 + a_3(s)\psi_{yy} + a_4(s)\psi_{yyyy}.
\end{equation}
The linear operator is
\begin{equation}
\mathcal{L}u = ((R_d-c)u)' + B_0\left[(f_d(fs)''u)'' + (f((f_ds + f)u)'')''\right]
\end{equation}
and the coefficients are
\begin{align}
a_1(s) &= s'  \label{eq:a1}\\
a_2(s) &= \half (R_d s)' + B_0\left[\half (f_ds(fs)'')'' + (f(fs)'')'' + (f(\half f_ds^2+fs)'')'' \right] \\\
a_3(s) &=  B_0\left[(f(fs)'')' + (f(fs)')'' \right]  \\\
a_4(s) &= B_0(f^2s)'  \label{eq:a4}
\end{align}

\end{widetext}

\section{Boundary conditions for the undercompressive wave}\label{sec:bdy}

We elaborate on the boundary conditions used in the solvability condition in the paragraph following Eqn [{\bf 6}].
These were justified rigorously in \citet{bertozzi2001}, but we include a heuristic version of the argument here for completeness, treating all possible types of traveling wave solutions to [{\bf 1}].

A general one-dimensional traveling wave solution has the form $h_x = s(x-ct)$
and solves
\begin{equation}\label{eq:tw}
c(s-b_r) - (R(s)-R(b_r)) - B_0\left(\frac{1}{\sqrt{1+s^2}}\left(\frac{s'}{(1+s^2)^{3/2}}\right)'\right)' = 0
\end{equation}
 with boundary conditions $s(-\infty) = b_l$, $s(+\infty) = b_r$, where $b_l, b_r$ are parameters \cite{mhc12}. The wave speed is found by integrating from $-\infty$ to $+\infty$ to be $c = (R(b_r)-R(b_l)) / (b_r-b_l)$. There are three types of waves, characterized by the relation between $c$ and the speed $R'(b_{l(r)})$ of information propagation  on either side of the wave: 
 \begin{itemize}
 \item A \emph{compressive} wave has $R'(b_l) > c > R'(b_r)$, so that information propagates into the wave from both sides; 
\item An \emph{undercompressive} wave has $R'(b_l) < c > R'(b_r)$ or $R'(b_l) > c < R'(b_r)$, so that information propagates in on one side and away on the other, 
\item A \emph{doubly-undercompressive} wave has $R'(b_l) < c < R'(b_r)$, so that information propagates away on both sides.  
\end{itemize}

A useful way of identifying the type of wave is by the dimensions of the invariant manifolds at the endpoints \cite{bertozzi99, mhc12}. Since \eqref{eq:tw} is third-order, a traveling wave can be thought of as a trajectory $(s,s',s'')$ in $\mathbb{R}^3$ connecting point $(b_l,0,0)$ to point $(b_r,0,0)$. It must lie in the intersection of the unstable manifold at $\eta=-\infty$ (written $\mathcal{US}(-\infty)$) and the stable manifold at $\eta=+\infty$ (written $\mathcal{S}(\infty)$). Linearizing \eqref{eq:tw} and looking for exponentially growing modes $\propto e^{\mu\eta}$
shows that $\mu^3 \propto c-R'(b_{l(r)})$, so $\mathcal{US}(-\infty)$  is two-dimensional when $c<R'(b_l)$ and one-dimensional otherwise, and $\mathcal{S}(\infty)$  is two-dimensional when $c> R'(b_r)$ and one-dimensional otherwise. Therefore a compressive wave occurs when the two invariant manifolds are two-dimensional, an undercompressive wave when one is one-dimensional and the other is two-dimensional, and a doubly-undercompressive wave when both are one-dimensional. 

To determine the boundary conditions for the equation $\mathcal{L}^*\pi = 0$,
we analyze the left and right eigenfunctions of the linear operator $\mathcal{L}$. 
We have that $\mathcal{L}s' = 0$, since this is simply the linearization of \eqref{eq:tw}.
Therefore 0 is an eigenvalue of $\mathcal{L}$ with right eigenfunction $s'$, so there is a corresponding left eigenfunction $\pi$ such that $\langle \pi, s'\rangle = 1$. To find the boundary conditions that make this normalization possible, 
we perturb the operator as $\mathcal{L}_\delta = \mathcal{L} + \delta\mathcal{L}_1$ and suppose it is analytic at the origin, so that the eigenvalues and eigenfunctions also have a perturbation expansion as  \footnote{Technically this operator as defined is not analytic at 0, but it can be made analytic by an appropriate change of variables; see \citet{bertozzi2001}.} 
\begin{align}\label{eq:delta}
\phi_\delta &= \phi + \delta\phi_1 + \delta^2\phi_3+ \ldots,\nonumber \\ 
\pi_\delta &= \pi + \delta\pi_1 +  \delta^2\pi_2 +\ldots, \nonumber\\
\lambda_\delta &= 0 + \delta\lambda_1 + \delta^2\lambda_2 +\ldots . 
\end{align}
Here $\phi_\delta$ is the right eigenfunction of $\mathcal{L}_\delta$, $\pi_\delta$ is the left eigenfunction, and $\lambda_\delta$ is the eigenvalue.
$\mathcal{L^*}$ will have the same leading-order boundary conditions as $\mathcal{L}^*_\delta$, which follow by considering the growth of $\phi_\delta, \pi_\delta$ near $\pm\infty$ and the condition $\langle \phi_\delta,\pi_\delta\rangle <\infty$.
Consider each case in turn:
 \begin{itemize}
 \item \emph{Compressive}: Then $\mathcal{US}(-\infty)$, $\mathcal{S}(\infty)$ are both two-dimensional, so generically they still intersect in a trajectory when perturbed. Therefore $\phi_\delta$ decays exponentially on both sides, so $\pi_\delta$ can grow on both sides. Requiring it to be bounded implies  $|\pi(-\infty)|, |\pi(\infty)| < const$.  
 
In this case the solution to $\mathcal{L}^*\pi=0$ is $\pi=const$; this is what one typically expects, for example for periodic traveling waves \cite[e.g.][]{shraiman86,bernoff88}. The constants in [{\bf 6}]
are computed analytically as 
\begin{align}
\bar{c}_1 &= (b_r-b_l), \nonumber\\
\bar{c}_2 &= R_d(b_r)b_r - R_d(b_l) b_l, \nonumber\\
\bar{c}_3 &= 0, \nonumber\\
\bar{c}_4 &= f^2(b_r)b_r - f^2(b_l)b_l ,
\end{align}
where $\bar{c}_i = \langle \pi, a_i(s)\rangle$.
 
 \item \emph{Undercompressive}: If one of the invariant manifolds is one-dimensional, then generically a perturbation to the equation will destroy the intersection. If $\mathcal{US}(-\infty)$ is one-dimensional, then $\phi_\delta$ will grow exponentially on the left, so $\pi_\delta$ must decay exponentially on the left in order to satisfy $\langle \phi, \pi \rangle <\infty$. Requiring it to be bounded implies the boundary conditions $\pi(-\infty) = 0$, $\pi(\infty) = const$; the constant can be chosen without loss of generality to be 1. 
 
 \item \emph{Doubly-undercompressive}:  A similar discussion to the undercompressive case implies the boundary conditions are $\pi(-\infty) = 0$, $\pi(\infty) = 0$.
 
 \end{itemize}

\section{Which terms to include in Theory 2?}\label{sec:terms}

Here we justify including a fourth-order term $c_4\psi_{yyyy}$ in [{\bf 6}].

The multiscale expansion to $O(\eps)$ would include only the terms $c_2\psi_y^2$, $c_3\psi_{yy}$. We have found that the viscous term  $c_3\psi_{yy}$ is sometimes very small -- for a compressive wave $c_3=0$ (this is explained in section \ref{sec:bdy}), and for an undercompressive wave we have found that by varying $R(b)$ the magnitude of $c_3$ can sometimes be very small. In addition there is no guarantee that the sign is positive. 
With no viscous nor fourth-order term, there is no smoothing mechanism: [{\bf 6}]
is a Hamilton-Jacobi equation that advects $\psi_{y}$ while causing it to sharpen until its derivatives blow up. Therefore smaller length scales are created and terms from the second-order equation will become as important as those in the first-order equation in the regions of high gradient.

Which terms will be the first to become large?
To answer this, we proceed as in a boundary-layer analysis and seek the largest length scale $L(\epsilon)$ such that at least one higher-order term becomes $O(\epsilon)$ under the scaling $\bar{y} = (L(\epsilon))^{-1}y$. In introducing this new scale we 
assume that $\psi_{y}$ maintains its original magnitude, i.e. $\psi_{y}\sim O(\epsilon^{1/2})$, but that its derivatives can become large due to the Hamilton-Jacobi dynamics. We look at the terms in the $O(\epsilon^2)$ equation, which has the form \footnote{Eqn \eqref{eq:O2}  is the expansion using a linearized 4th-order term $\propto \lapl^2 h$; the expansion for the full nonlinear smoothing is similar but has more terms.} 
 \begin{widetext}
\begin{multline}\label{eq:O2}
v_t + \mathcal{L} v
=   (\cdot) \psi_{yyyy} 
 + (\cdot)\psi_{y}\psi_{yyy} + (\cdot)\psi_{yy}^2 + (\cdot)\psi_{y}^2\psi_{yy} + (\cdot)\psi_{y}^4 
 + \big((\cdot)u\big)_\eta\psi_{y}^2 +  (\cdot)u_{\eta\eta\eta}\psi_{yy} \\ + (\cdot)u_{\eta\eta\eta y}\psi_{y} 
 + (\cdot)u_{\eta\eta\eta\eta}\psi_{y}^2  
 + u_{\eta}\psi_{\bar{t}} + \big((\cdot)\int_\eta u_y\big)_\eta\psi_{y} 
 + \big((\cdot) u^2\big)_\eta + (\cdot)u_{\eta\eta yy}    .
\end{multline}
\end{widetext}
Here $v$ is the $O(\epsilon^2)$ perturbation to $h_x$,  and $(\cdot)$ represents some function of $s(\eta)$. 
Somewhat surprisingly, the right-hand side includes terms proportional to $u_{(\cdot)}$. For a compressive wave, the coefficients of these terms after integrating over the fast variables are $c_i=0$, but for a non-compressive wave the perturbations evolve nonlinearly in general \cite{liuzumbrun95}.

By applying the new scaling to each of the terms, we find that when $L(\epsilon) = \epsilon^{1/6}$ then $\psi_{yyyy} \to \psi_{\bar{y}y\bar{y}\bar{y}\bar{y}} \sim \epsilon^{1/2 + 3\cdot 1/6} = \epsilon$ but that all other terms are higher-order. Therefore we include this term and obtain a well-posed curve evolution equation: smaller scales can be created by the Hamilton-Jacobi dynamics, but these are subsequently suppressed by the fourth-order smoothing before other terms become important.

\end{document}

%% file: commands_may11.tex
\usepackage{amsmath,amssymb, verbatim}

\newcommand{\dd}[2]{\frac{d #1}{d #2}}

\newcommand{\pp}[2]{\frac{\partial #1}{\partial #2}}

\newcommand{\oneover}[1]{\frac{1}{#1}}

\newcommand{\half}{\frac{1}{2}}

\newcommand{\grad}{\nabla}
\newcommand{\lapl}{\Delta}

\renewcommand{\div}{\nabla \cdot}







%% file: CurveEvolution_arxiv_2016_resub.bbl
\begin{thebibliography}{47}%
\makeatletter
\providecommand \@ifxundefined [1]{%
 \@ifx{#1\undefined}
}%
\providecommand \@ifnum [1]{%
 \ifnum #1\expandafter \@firstoftwo
 \else \expandafter \@secondoftwo
 \fi
}%
\providecommand \@ifx [1]{%
 \ifx #1\expandafter \@firstoftwo
 \else \expandafter \@secondoftwo
 \fi
}%
\providecommand \natexlab [1]{#1}%
\providecommand \enquote  [1]{``#1''}%
\providecommand \bibnamefont  [1]{#1}%
\providecommand \bibfnamefont [1]{#1}%
\providecommand \citenamefont [1]{#1}%
\providecommand \href@noop [0]{\@secondoftwo}%
\providecommand \href [0]{\begingroup \@sanitize@url \@href}%
\providecommand \@href[1]{\@@startlink{#1}\@@href}%
\providecommand \@@href[1]{\endgroup#1\@@endlink}%
\providecommand \@sanitize@url [0]{\catcode `\\12\catcode `\$12\catcode
  `\&12\catcode `\#12\catcode `\^12\catcode `\_12\catcode `\%12\relax}%
\providecommand \@@startlink[1]{}%
\providecommand \@@endlink[0]{}%
\providecommand \url  [0]{\begingroup\@sanitize@url \@url }%
\providecommand \@url [1]{\endgroup\@href {#1}{\urlprefix }}%
\providecommand \urlprefix  [0]{URL }%
\providecommand \Eprint [0]{\href }%
\providecommand \doibase [0]{http://dx.doi.org/}%
\providecommand \selectlanguage [0]{\@gobble}%
\providecommand \bibinfo  [0]{\@secondoftwo}%
\providecommand \bibfield  [0]{\@secondoftwo}%
\providecommand \translation [1]{[#1]}%
\providecommand \BibitemOpen [0]{}%
\providecommand \bibitemStop [0]{}%
\providecommand \bibitemNoStop [0]{.\EOS\space}%
\providecommand \EOS [0]{\spacefactor3000\relax}%
\providecommand \BibitemShut  [1]{\csname bibitem#1\endcsname}%
\let\auto@bib@innerbib\@empty
\bibitem [{\citenamefont {Vasile}\ \emph {et~al.}(1997)\citenamefont {Vasile},
  \citenamefont {Niu}, \citenamefont {Nassar}, \citenamefont {Zhang},\ and\
  \citenamefont {Liu}}]{vasile97}%
  \BibitemOpen
  \bibfield  {author} {\bibinfo {author} {\bibfnamefont {M.~J.}\ \bibnamefont
  {Vasile}}, \bibinfo {author} {\bibfnamefont {Z.}~\bibnamefont {Niu}},
  \bibinfo {author} {\bibfnamefont {R.}~\bibnamefont {Nassar}}, \bibinfo
  {author} {\bibfnamefont {W.}~\bibnamefont {Zhang}}, \ and\ \bibinfo {author}
  {\bibfnamefont {S.}~\bibnamefont {Liu}},\ }\href@noop {} {\bibfield
  {journal} {\bibinfo  {journal} {J. Vac. Sci. Technol. B}\ }\textbf {\bibinfo
  {volume} {15}},\ \bibinfo {pages} {2350} (\bibinfo {year}
  {1997})}\BibitemShut {NoStop}%
\bibitem [{\citenamefont {Adams}\ \emph {et~al.}(2003)\citenamefont {Adams},
  \citenamefont {Vasile}, \citenamefont {Mayer},\ and\ \citenamefont
  {Hodges}}]{adams03}%
  \BibitemOpen
  \bibfield  {author} {\bibinfo {author} {\bibfnamefont {D.}~\bibnamefont
  {Adams}}, \bibinfo {author} {\bibfnamefont {M.}~\bibnamefont {Vasile}},
  \bibinfo {author} {\bibfnamefont {T.}~\bibnamefont {Mayer}}, \ and\ \bibinfo
  {author} {\bibfnamefont {V.}~\bibnamefont {Hodges}},\ }\href@noop {}
  {\bibfield  {journal} {\bibinfo  {journal} {J. Vac. Sci. Technol. B}\
  }\textbf {\bibinfo {volume} {21}},\ \bibinfo {pages} {2334} (\bibinfo {year}
  {2003})}\BibitemShut {NoStop}%
\bibitem [{\citenamefont {Li}\ \emph {et~al.}(2001)\citenamefont {Li},
  \citenamefont {Stein}, \citenamefont {McMullan}, \citenamefont {Branton},
  \citenamefont {Aziz},\ and\ \citenamefont {Golovchenko}}]{li01}%
  \BibitemOpen
  \bibfield  {author} {\bibinfo {author} {\bibfnamefont {J.}~\bibnamefont
  {Li}}, \bibinfo {author} {\bibfnamefont {D.}~\bibnamefont {Stein}}, \bibinfo
  {author} {\bibfnamefont {C.}~\bibnamefont {McMullan}}, \bibinfo {author}
  {\bibfnamefont {D.}~\bibnamefont {Branton}}, \bibinfo {author} {\bibfnamefont
  {M.~J.}\ \bibnamefont {Aziz}}, \ and\ \bibinfo {author} {\bibfnamefont
  {J.~A.}\ \bibnamefont {Golovchenko}},\ }\href@noop {} {\bibfield  {journal}
  {\bibinfo  {journal} {Nature}\ }\textbf {\bibinfo {volume} {412}},\ \bibinfo
  {pages} {166} (\bibinfo {year} {2001})}\BibitemShut {NoStop}%
\bibitem [{\citenamefont {Stein}\ \emph {et~al.}(2002)\citenamefont {Stein},
  \citenamefont {Li},\ and\ \citenamefont {Golovchenko}}]{stein02}%
  \BibitemOpen
  \bibfield  {author} {\bibinfo {author} {\bibfnamefont {D.}~\bibnamefont
  {Stein}}, \bibinfo {author} {\bibfnamefont {J.}~\bibnamefont {Li}}, \ and\
  \bibinfo {author} {\bibfnamefont {J.~A.}\ \bibnamefont {Golovchenko}},\
  }\href@noop {} {\bibfield  {journal} {\bibinfo  {journal} {Phys. Rev. Lett.}\
  }\textbf {\bibinfo {volume} {89}},\ \bibinfo {pages} {276106} (\bibinfo
  {year} {2002})}\BibitemShut {NoStop}%
\bibitem [{\citenamefont {Sigmund}(1969)}]{sigmund69}%
  \BibitemOpen
  \bibfield  {author} {\bibinfo {author} {\bibfnamefont {P.}~\bibnamefont
  {Sigmund}},\ }\href@noop {} {\bibfield  {journal} {\bibinfo  {journal} {Phys.
  Rev.}\ }\textbf {\bibinfo {volume} {184}},\ \bibinfo {pages} {383} (\bibinfo
  {year} {1969})}\BibitemShut {NoStop}%
\bibitem [{\citenamefont {Sigmund}(1973)}]{sigmund73}%
  \BibitemOpen
  \bibfield  {author} {\bibinfo {author} {\bibfnamefont {P.}~\bibnamefont
  {Sigmund}},\ }\href@noop {} {\bibfield  {journal} {\bibinfo  {journal} {J.
  Mater. Sci.}\ }\textbf {\bibinfo {volume} {8}},\ \bibinfo {pages} {1545}
  (\bibinfo {year} {1973})}\BibitemShut {NoStop}%
\bibitem [{\citenamefont {Bradley}\ and\ \citenamefont
  {Harper}(1988)}]{bradley88}%
  \BibitemOpen
  \bibfield  {author} {\bibinfo {author} {\bibfnamefont {R.~M.}\ \bibnamefont
  {Bradley}}\ and\ \bibinfo {author} {\bibfnamefont {J.~M.~E.}\ \bibnamefont
  {Harper}},\ }\href@noop {} {\bibfield  {journal} {\bibinfo  {journal} {J.
  Vac. Sci. Technol. A}\ }\textbf {\bibinfo {volume} {6}},\ \bibinfo {pages}
  {2390} (\bibinfo {year} {1988})}\BibitemShut {NoStop}%
\bibitem [{\citenamefont {Chan}\ and\ \citenamefont {Chason}(2007)}]{chason}%
  \BibitemOpen
  \bibfield  {author} {\bibinfo {author} {\bibfnamefont {W.~L.}\ \bibnamefont
  {Chan}}\ and\ \bibinfo {author} {\bibfnamefont {E.}~\bibnamefont {Chason}},\
  }\href@noop {} {\bibfield  {journal} {\bibinfo  {journal} {J. Appl. Phys.}\
  }\textbf {\bibinfo {volume} {101}},\ \bibinfo {pages} {121301} (\bibinfo
  {year} {2007})}\BibitemShut {NoStop}%
\bibitem [{\citenamefont {Mu{\~n}oz-Garcia}\ \emph {et~al.}(2009)\citenamefont
  {Mu{\~n}oz-Garcia}, \citenamefont {V{\'a}zquez}, \citenamefont {Cuerno},
  \citenamefont {S{\'a}nchez-Garc{\'i}a}, \citenamefont {Castro},\ and\
  \citenamefont {Gago}}]{munoz2009}%
  \BibitemOpen
  \bibfield  {author} {\bibinfo {author} {\bibfnamefont {J.}~\bibnamefont
  {Mu{\~n}oz-Garcia}}, \bibinfo {author} {\bibfnamefont {L.}~\bibnamefont
  {V{\'a}zquez}}, \bibinfo {author} {\bibfnamefont {R.}~\bibnamefont {Cuerno}},
  \bibinfo {author} {\bibfnamefont {J.~A.}\ \bibnamefont
  {S{\'a}nchez-Garc{\'i}a}}, \bibinfo {author} {\bibfnamefont {M.}~\bibnamefont
  {Castro}}, \ and\ \bibinfo {author} {\bibfnamefont {R.}~\bibnamefont
  {Gago}},\ }\href@noop {} {\bibfield  {journal} {\bibinfo  {journal} {Toward
  Functional Nanomaterials}\ }\textbf {\bibinfo {volume} {Lecture Notes}},\
  \bibinfo {pages} {323} (\bibinfo {year} {2009})}\BibitemShut {NoStop}%
\bibitem [{\citenamefont {Facsko}\ \emph {et~al.}(1999)\citenamefont {Facsko},
  \citenamefont {Dekorsy}, \citenamefont {Koerdt}, \citenamefont {Trappe},
  \citenamefont {Kurz}, \citenamefont {Vogt},\ and\ \citenamefont
  {Hartnagel}}]{facsko99}%
  \BibitemOpen
  \bibfield  {author} {\bibinfo {author} {\bibfnamefont {S.}~\bibnamefont
  {Facsko}}, \bibinfo {author} {\bibfnamefont {T.}~\bibnamefont {Dekorsy}},
  \bibinfo {author} {\bibfnamefont {C.}~\bibnamefont {Koerdt}}, \bibinfo
  {author} {\bibfnamefont {C.}~\bibnamefont {Trappe}}, \bibinfo {author}
  {\bibfnamefont {H.}~\bibnamefont {Kurz}}, \bibinfo {author} {\bibfnamefont
  {A.}~\bibnamefont {Vogt}}, \ and\ \bibinfo {author} {\bibfnamefont {H.~L.}\
  \bibnamefont {Hartnagel}},\ }\href@noop {} {\bibfield  {journal} {\bibinfo
  {journal} {Science}\ }\textbf {\bibinfo {volume} {285}},\ \bibinfo {pages}
  {1551} (\bibinfo {year} {1999})}\BibitemShut {NoStop}%
\bibitem [{\citenamefont {Frost}\ \emph {et~al.}(2000)\citenamefont {Frost},
  \citenamefont {Schindler},\ and\ \citenamefont {Bigl}}]{frost00}%
  \BibitemOpen
  \bibfield  {author} {\bibinfo {author} {\bibfnamefont {F.}~\bibnamefont
  {Frost}}, \bibinfo {author} {\bibfnamefont {A.}~\bibnamefont {Schindler}}, \
  and\ \bibinfo {author} {\bibfnamefont {F.}~\bibnamefont {Bigl}},\ }\href@noop
  {} {\bibfield  {journal} {\bibinfo  {journal} {Phys. Rev. Lett.}\ }\textbf
  {\bibinfo {volume} {85}},\ \bibinfo {pages} {4116} (\bibinfo {year}
  {2000})}\BibitemShut {NoStop}%
\bibitem [{\citenamefont {Cuenat}\ \emph {et~al.}(2005)\citenamefont {Cuenat},
  \citenamefont {George}, \citenamefont {Chang}, \citenamefont {Blakely},\ and\
  \citenamefont {Aziz}}]{cuenat05}%
  \BibitemOpen
  \bibfield  {author} {\bibinfo {author} {\bibfnamefont {A.}~\bibnamefont
  {Cuenat}}, \bibinfo {author} {\bibfnamefont {H.~B.}\ \bibnamefont {George}},
  \bibinfo {author} {\bibfnamefont {K.-C.}\ \bibnamefont {Chang}}, \bibinfo
  {author} {\bibfnamefont {J.}~\bibnamefont {Blakely}}, \ and\ \bibinfo
  {author} {\bibfnamefont {M.~J.}\ \bibnamefont {Aziz}},\ }\href@noop {}
  {\bibfield  {journal} {\bibinfo  {journal} {Adv. Mater.}\ }\textbf {\bibinfo
  {volume} {17}},\ \bibinfo {pages} {2845} (\bibinfo {year}
  {2005})}\BibitemShut {NoStop}%
\bibitem [{\citenamefont {Castro}\ \emph {et~al.}(2005)\citenamefont {Castro},
  \citenamefont {Cuerno}, \citenamefont {Vazquez},\ and\ \citenamefont
  {Gago}}]{castro2005}%
  \BibitemOpen
  \bibfield  {author} {\bibinfo {author} {\bibfnamefont {M.}~\bibnamefont
  {Castro}}, \bibinfo {author} {\bibfnamefont {R.}~\bibnamefont {Cuerno}},
  \bibinfo {author} {\bibfnamefont {L.}~\bibnamefont {Vazquez}}, \ and\
  \bibinfo {author} {\bibfnamefont {R.}~\bibnamefont {Gago}},\ }\href@noop {}
  {\bibfield  {journal} {\bibinfo  {journal} {Phys. Rev. Lett.}\ }\textbf
  {\bibinfo {volume} {84}},\ \bibinfo {pages} {016102} (\bibinfo {year}
  {2005})}\BibitemShut {NoStop}%
\bibitem [{\citenamefont {Wei}\ \emph {et~al.}(2008)\citenamefont {Wei},
  \citenamefont {Lian}, \citenamefont {Zhy}, \citenamefont {Li}, \citenamefont
  {Sun},\ and\ \citenamefont {Wang}}]{wei2008}%
  \BibitemOpen
  \bibfield  {author} {\bibinfo {author} {\bibfnamefont {Q.}~\bibnamefont
  {Wei}}, \bibinfo {author} {\bibfnamefont {J.}~\bibnamefont {Lian}}, \bibinfo
  {author} {\bibfnamefont {S.}~\bibnamefont {Zhy}}, \bibinfo {author}
  {\bibfnamefont {W.}~\bibnamefont {Li}}, \bibinfo {author} {\bibfnamefont
  {K.}~\bibnamefont {Sun}}, \ and\ \bibinfo {author} {\bibfnamefont
  {L.}~\bibnamefont {Wang}},\ }\href@noop {} {\bibfield  {journal} {\bibinfo
  {journal} {Chem. Phys. Lett.}\ }\textbf {\bibinfo {volume} {452}},\ \bibinfo
  {pages} {124} (\bibinfo {year} {2008})}\BibitemShut {NoStop}%
\bibitem [{\citenamefont {Ziberi}\ \emph {et~al.}(2008)\citenamefont {Ziberi},
  \citenamefont {Front}, \citenamefont {Tartz}, \citenamefont {Neumann},\ and\
  \citenamefont {Rauschenbach}}]{ziberi2008}%
  \BibitemOpen
  \bibfield  {author} {\bibinfo {author} {\bibfnamefont {B.}~\bibnamefont
  {Ziberi}}, \bibinfo {author} {\bibfnamefont {F.}~\bibnamefont {Front}},
  \bibinfo {author} {\bibfnamefont {M.}~\bibnamefont {Tartz}}, \bibinfo
  {author} {\bibfnamefont {H.}~\bibnamefont {Neumann}}, \ and\ \bibinfo
  {author} {\bibfnamefont {B.}~\bibnamefont {Rauschenbach}},\ }\href@noop {}
  {\bibfield  {journal} {\bibinfo  {journal} {Appl. Phys. Lett.}\ }\textbf
  {\bibinfo {volume} {92}},\ \bibinfo {pages} {063102} (\bibinfo {year}
  {2008})}\BibitemShut {NoStop}%
\bibitem [{\citenamefont {Mu{\~n}oz-Garcia}\ \emph {et~al.}(2010)\citenamefont
  {Mu{\~n}oz-Garcia}, \citenamefont {Gago}, \citenamefont {V{\'a}zquez},
  \citenamefont {Sanchez-Garcia},\ and\ \citenamefont {Cuerno}}]{munoz2010}%
  \BibitemOpen
  \bibfield  {author} {\bibinfo {author} {\bibfnamefont {J.}~\bibnamefont
  {Mu{\~n}oz-Garcia}}, \bibinfo {author} {\bibfnamefont {R.}~\bibnamefont
  {Gago}}, \bibinfo {author} {\bibfnamefont {L.}~\bibnamefont {V{\'a}zquez}},
  \bibinfo {author} {\bibfnamefont {J.~A.}\ \bibnamefont {Sanchez-Garcia}}, \
  and\ \bibinfo {author} {\bibfnamefont {R.}~\bibnamefont {Cuerno}},\
  }\href@noop {} {\bibfield  {journal} {\bibinfo  {journal} {Phys. Rev. Lett.}\
  }\textbf {\bibinfo {volume} {104}},\ \bibinfo {pages} {026101} (\bibinfo
  {year} {2010})}\BibitemShut {NoStop}%
\bibitem [{\citenamefont {Bradley}\ and\ \citenamefont
  {Shipman}(2010)}]{bradley2010}%
  \BibitemOpen
  \bibfield  {author} {\bibinfo {author} {\bibfnamefont {R.}~\bibnamefont
  {Bradley}}\ and\ \bibinfo {author} {\bibfnamefont {P.~D.}\ \bibnamefont
  {Shipman}},\ }\href@noop {} {\bibfield  {journal} {\bibinfo  {journal} {Phys.
  Rev. Lett.}\ }\textbf {\bibinfo {volume} {105}},\ \bibinfo {pages} {145501}
  (\bibinfo {year} {2010})}\BibitemShut {NoStop}%
\bibitem [{\citenamefont {Smythe}\ \emph {et~al.}(2007)\citenamefont {Smythe},
  \citenamefont {Cubukcu},\ and\ \citenamefont {Capasso}}]{smythe2007}%
  \BibitemOpen
  \bibfield  {author} {\bibinfo {author} {\bibfnamefont {E.}~\bibnamefont
  {Smythe}}, \bibinfo {author} {\bibfnamefont {E.}~\bibnamefont {Cubukcu}}, \
  and\ \bibinfo {author} {\bibfnamefont {F.}~\bibnamefont {Capasso}},\
  }\href@noop {} {\bibfield  {journal} {\bibinfo  {journal} {Optics Express}\
  }\textbf {\bibinfo {volume} {15}},\ \bibinfo {pages} {7439} (\bibinfo {year}
  {2007})}\BibitemShut {NoStop}%
\bibitem [{\citenamefont {Rockstuhl}\ \emph {et~al.}(2006)\citenamefont
  {Rockstuhl}, \citenamefont {Lederer}, \citenamefont {Etrich}, \citenamefont
  {Zentgraf}, \citenamefont {Kuhl},\ and\ \citenamefont
  {Giessen}}]{rockstuhl2006}%
  \BibitemOpen
  \bibfield  {author} {\bibinfo {author} {\bibfnamefont {C.}~\bibnamefont
  {Rockstuhl}}, \bibinfo {author} {\bibfnamefont {F.}~\bibnamefont {Lederer}},
  \bibinfo {author} {\bibfnamefont {C.}~\bibnamefont {Etrich}}, \bibinfo
  {author} {\bibfnamefont {T.}~\bibnamefont {Zentgraf}}, \bibinfo {author}
  {\bibfnamefont {J.}~\bibnamefont {Kuhl}}, \ and\ \bibinfo {author}
  {\bibfnamefont {H.}~\bibnamefont {Giessen}},\ }\href@noop {} {\bibfield
  {journal} {\bibinfo  {journal} {Optics Express}\ }\textbf {\bibinfo {volume}
  {14}},\ \bibinfo {pages} {8827} (\bibinfo {year} {2006})}\BibitemShut
  {NoStop}%
\bibitem [{\citenamefont {Chen}\ \emph {et~al.}(2005)\citenamefont {Chen},
  \citenamefont {Urquidez}, \citenamefont {Ichim}, \citenamefont {Rodriguez},
  \citenamefont {Brenner},\ and\ \citenamefont {Aziz}}]{chen2005}%
  \BibitemOpen
  \bibfield  {author} {\bibinfo {author} {\bibfnamefont {H.}~\bibnamefont
  {Chen}}, \bibinfo {author} {\bibfnamefont {O.}~\bibnamefont {Urquidez}},
  \bibinfo {author} {\bibfnamefont {S.}~\bibnamefont {Ichim}}, \bibinfo
  {author} {\bibfnamefont {L.}~\bibnamefont {Rodriguez}}, \bibinfo {author}
  {\bibfnamefont {M.}~\bibnamefont {Brenner}}, \ and\ \bibinfo {author}
  {\bibfnamefont {M.}~\bibnamefont {Aziz}},\ }\href {\doibase DOI
  10.1126/science.1117219} {\bibfield  {journal} {\bibinfo  {journal}
  {Science}\ }\textbf {\bibinfo {volume} {310}},\ \bibinfo {pages} {294}
  (\bibinfo {year} {2005})}\BibitemShut {NoStop}%
\bibitem [{\citenamefont {Miller}\ \emph {et~al.}(2005)\citenamefont {Miller},
  \citenamefont {Russell},\ and\ \citenamefont {Thompson}}]{miller2005}%
  \BibitemOpen
  \bibfield  {author} {\bibinfo {author} {\bibfnamefont {M.}~\bibnamefont
  {Miller}}, \bibinfo {author} {\bibfnamefont {K.}~\bibnamefont {Russell}}, \
  and\ \bibinfo {author} {\bibfnamefont {G.}~\bibnamefont {Thompson}},\
  }\href@noop {} {\bibfield  {journal} {\bibinfo  {journal} {Ultramicroscopy}\
  }\textbf {\bibinfo {volume} {102}},\ \bibinfo {pages} {287} (\bibinfo {year}
  {2005})}\BibitemShut {NoStop}%
\bibitem [{\citenamefont {Loi}\ \emph {et~al.}(2013)\citenamefont {Loi},
  \citenamefont {Gault}, \citenamefont {Ringer}, \citenamefont {Larson},\ and\
  \citenamefont {Geiser}}]{loi2013}%
  \BibitemOpen
  \bibfield  {author} {\bibinfo {author} {\bibfnamefont {S.}~\bibnamefont
  {Loi}}, \bibinfo {author} {\bibfnamefont {B.}~\bibnamefont {Gault}}, \bibinfo
  {author} {\bibfnamefont {S.}~\bibnamefont {Ringer}}, \bibinfo {author}
  {\bibfnamefont {D.}~\bibnamefont {Larson}}, \ and\ \bibinfo {author}
  {\bibfnamefont {B.}~\bibnamefont {Geiser}},\ }\href@noop {} {\bibfield
  {journal} {\bibinfo  {journal} {Ultramicroscopy}\ }\textbf {\bibinfo {volume}
  {132}},\ \bibinfo {pages} {107} (\bibinfo {year} {2013})}\BibitemShut
  {NoStop}%
\bibitem [{\citenamefont {Holmes-Cerfon}\ \emph
  {et~al.}(2012{\natexlab{a}})\citenamefont {Holmes-Cerfon}, \citenamefont
  {Aziz},\ and\ \citenamefont {Brenner}}]{mhc12}%
  \BibitemOpen
  \bibfield  {author} {\bibinfo {author} {\bibfnamefont {M.}~\bibnamefont
  {Holmes-Cerfon}}, \bibinfo {author} {\bibfnamefont {M.}~\bibnamefont {Aziz}},
  \ and\ \bibinfo {author} {\bibfnamefont {M.~P.}\ \bibnamefont {Brenner}},\
  }\href@noop {} {\bibfield  {journal} {\bibinfo  {journal} {Phy. Rev. B}\
  }\textbf {\bibinfo {volume} {85}},\ \bibinfo {pages} {165441} (\bibinfo
  {year} {2012}{\natexlab{a}})}\BibitemShut {NoStop}%
\bibitem [{\citenamefont {Holmes-Cerfon}\ \emph
  {et~al.}(2012{\natexlab{b}})\citenamefont {Holmes-Cerfon}, \citenamefont
  {Zhou}, \citenamefont {Bertozzi}, \citenamefont {Brenner},\ and\
  \citenamefont {Aziz}}]{mhc12b}%
  \BibitemOpen
  \bibfield  {author} {\bibinfo {author} {\bibfnamefont {M.}~\bibnamefont
  {Holmes-Cerfon}}, \bibinfo {author} {\bibfnamefont {W.}~\bibnamefont {Zhou}},
  \bibinfo {author} {\bibfnamefont {A.~L.}\ \bibnamefont {Bertozzi}}, \bibinfo
  {author} {\bibfnamefont {M.~P.}\ \bibnamefont {Brenner}}, \ and\ \bibinfo
  {author} {\bibfnamefont {M.~J.}\ \bibnamefont {Aziz}},\ }\href@noop {}
  {\bibfield  {journal} {\bibinfo  {journal} {Appl. Phys. Lett.}\ }\textbf
  {\bibinfo {volume} {101}},\ \bibinfo {pages} {143109} (\bibinfo {year}
  {2012}{\natexlab{b}})}\BibitemShut {NoStop}%
\bibitem [{\citenamefont {Osher}\ and\ \citenamefont
  {Sethian}(1979)}]{osher79}%
  \BibitemOpen
  \bibfield  {author} {\bibinfo {author} {\bibfnamefont {S.}~\bibnamefont
  {Osher}}\ and\ \bibinfo {author} {\bibfnamefont {J.}~\bibnamefont
  {Sethian}},\ }\href@noop {} {\bibfield  {journal} {\bibinfo  {journal} {J.
  Comp. Phys.}\ }\textbf {\bibinfo {volume} {1}},\ \bibinfo {pages} {12}
  (\bibinfo {year} {1979})}\BibitemShut {NoStop}%
\bibitem [{\citenamefont {Sethian}(1996)}]{sethian96}%
  \BibitemOpen
  \bibfield  {author} {\bibinfo {author} {\bibfnamefont {J.}~\bibnamefont
  {Sethian}},\ }\href@noop {} {\emph {\bibinfo {title} {Level Set Methods:
  Evolving Interfaces in Geometry, Fluid Mechanics, Computer Vision and
  Material Sciences}}}\ (\bibinfo  {publisher} {Cambridge University Press},\
  \bibinfo {year} {1996})\BibitemShut {NoStop}%
\bibitem [{\citenamefont {Madi}\ \emph {et~al.}(2011)\citenamefont {Madi},
  \citenamefont {Anzenberg}, \citenamefont {{Ludwig Jr.}},\ and\ \citenamefont
  {Aziz}}]{madi2011}%
  \BibitemOpen
  \bibfield  {author} {\bibinfo {author} {\bibfnamefont {C.~S.}\ \bibnamefont
  {Madi}}, \bibinfo {author} {\bibfnamefont {E.}~\bibnamefont {Anzenberg}},
  \bibinfo {author} {\bibfnamefont {K.~F.}\ \bibnamefont {{Ludwig Jr.}}}, \
  and\ \bibinfo {author} {\bibfnamefont {M.~J.}\ \bibnamefont {Aziz}},\
  }\href@noop {} {\bibfield  {journal} {\bibinfo  {journal} {Phys. Rev. Lett.}\
  }\textbf {\bibinfo {volume} {106}},\ \bibinfo {pages} {066101} (\bibinfo
  {year} {2011})}\BibitemShut {NoStop}%
\bibitem [{\citenamefont {Norris}\ \emph {et~al.}(2011)\citenamefont {Norris},
  \citenamefont {Samela}, \citenamefont {Bukonte}, \citenamefont {Backman},
  \citenamefont {Djurabekova}, \citenamefont {Nordlund}, \citenamefont {Madi},
  \citenamefont {Brenner},\ and\ \citenamefont {Aziz}}]{norris2011}%
  \BibitemOpen
  \bibfield  {author} {\bibinfo {author} {\bibfnamefont {S.}~\bibnamefont
  {Norris}}, \bibinfo {author} {\bibfnamefont {J.}~\bibnamefont {Samela}},
  \bibinfo {author} {\bibfnamefont {L.}~\bibnamefont {Bukonte}}, \bibinfo
  {author} {\bibfnamefont {M.}~\bibnamefont {Backman}}, \bibinfo {author}
  {\bibfnamefont {F.}~\bibnamefont {Djurabekova}}, \bibinfo {author}
  {\bibfnamefont {K.}~\bibnamefont {Nordlund}}, \bibinfo {author}
  {\bibfnamefont {C.~S.}\ \bibnamefont {Madi}}, \bibinfo {author}
  {\bibfnamefont {M.~P.}\ \bibnamefont {Brenner}}, \ and\ \bibinfo {author}
  {\bibfnamefont {M.~J.}\ \bibnamefont {Aziz}},\ }\href@noop {} {\bibfield
  {journal} {\bibinfo  {journal} {Nature Communications}\ }\textbf {\bibinfo
  {volume} {2}},\ \bibinfo {pages} {276} (\bibinfo {year} {2011})}\BibitemShut
  {NoStop}%
\bibitem [{\citenamefont {Aziz}(2006)}]{aziz2006}%
  \BibitemOpen
  \bibfield  {author} {\bibinfo {author} {\bibfnamefont {M.~J.}\ \bibnamefont
  {Aziz}},\ }\bibfield  {booktitle} {\emph {\bibinfo {booktitle}
  {Matematisk-Fysiske Meddelelser / udg. af Det Kongelige Danske Videnskabernes
  Selskab}},\ }\href@noop {} {\bibfield  {journal} {\bibinfo  {journal}
  {Matematisk-Fysiske Meddelelser / udg. af Det Kongelige Danske Videnskabernes
  Selskab}\ }\bibinfo {series} {Ion'06 Proceedings},\ \textbf {\bibinfo
  {volume} {52}},\ \bibinfo {pages} {187} (\bibinfo {year} {2006})}\BibitemShut
  {NoStop}%
\bibitem [{\citenamefont {Mullins}(1959)}]{mullins1959}%
  \BibitemOpen
  \bibfield  {author} {\bibinfo {author} {\bibfnamefont {W.}~\bibnamefont
  {Mullins}},\ }\href@noop {} {\bibfield  {journal} {\bibinfo  {journal} {J.
  Appl. Phys.}\ }\textbf {\bibinfo {volume} {30}},\ \bibinfo {pages} {77}
  (\bibinfo {year} {1959})}\BibitemShut {NoStop}%
\bibitem [{\citenamefont {Herring}(1950)}]{herring1950}%
  \BibitemOpen
  \bibfield  {author} {\bibinfo {author} {\bibfnamefont {C.}~\bibnamefont
  {Herring}},\ }\href@noop {} {\bibfield  {journal} {\bibinfo  {journal} {J.
  Appl. Phys.}\ }\textbf {\bibinfo {volume} {21}},\ \bibinfo {pages} {301}
  (\bibinfo {year} {1950})}\BibitemShut {NoStop}%
\bibitem [{\citenamefont {Umbach}\ \emph {et~al.}(2001)\citenamefont {Umbach},
  \citenamefont {Headrick},\ and\ \citenamefont {Chang}}]{umbach2001}%
  \BibitemOpen
  \bibfield  {author} {\bibinfo {author} {\bibfnamefont {C.}~\bibnamefont
  {Umbach}}, \bibinfo {author} {\bibfnamefont {R.}~\bibnamefont {Headrick}}, \
  and\ \bibinfo {author} {\bibfnamefont {K.}~\bibnamefont {Chang}},\
  }\href@noop {} {\bibfield  {journal} {\bibinfo  {journal} {Phys. Rev. Lett.}\
  }\textbf {\bibinfo {volume} {87}},\ \bibinfo {pages} {246104} (\bibinfo
  {year} {2001})}\BibitemShut {NoStop}%
\bibitem [{\citenamefont {Davidovitch}\ \emph {et~al.}(2009)\citenamefont
  {Davidovitch}, \citenamefont {Aziz},\ and\ \citenamefont
  {Brenner}}]{davidovitch2009}%
  \BibitemOpen
  \bibfield  {author} {\bibinfo {author} {\bibfnamefont {B.}~\bibnamefont
  {Davidovitch}}, \bibinfo {author} {\bibfnamefont {M.~J.}\ \bibnamefont
  {Aziz}}, \ and\ \bibinfo {author} {\bibfnamefont {M.~P.}\ \bibnamefont
  {Brenner}},\ }\href@noop {} {\bibfield  {journal} {\bibinfo  {journal} {J.
  Phys.: Condens. Matter}\ }\textbf {\bibinfo {volume} {21}},\ \bibinfo {pages}
  {22} (\bibinfo {year} {2009})}\BibitemShut {NoStop}%
\bibitem [{\citenamefont {Bertozzi}\ \emph {et~al.}(1999)\citenamefont
  {Bertozzi}, \citenamefont {Munch},\ and\ \citenamefont
  {Shearer}}]{bertozzi99}%
  \BibitemOpen
  \bibfield  {author} {\bibinfo {author} {\bibfnamefont {A.}~\bibnamefont
  {Bertozzi}}, \bibinfo {author} {\bibfnamefont {A.}~\bibnamefont {Munch}}, \
  and\ \bibinfo {author} {\bibfnamefont {M.}~\bibnamefont {Shearer}},\
  }\href@noop {} {\bibfield  {journal} {\bibinfo  {journal} {Physica D}\
  }\textbf {\bibinfo {volume} {134}},\ \bibinfo {pages} {431} (\bibinfo {year}
  {1999})}\BibitemShut {NoStop}%
\bibitem [{\citenamefont {Bertozzi}\ \emph {et~al.}(2001)\citenamefont
  {Bertozzi}, \citenamefont {Munch}, \citenamefont {Shearer},\ and\
  \citenamefont {Zumbrun}}]{bertozzi2001}%
  \BibitemOpen
  \bibfield  {author} {\bibinfo {author} {\bibfnamefont {A.}~\bibnamefont
  {Bertozzi}}, \bibinfo {author} {\bibfnamefont {A.}~\bibnamefont {Munch}},
  \bibinfo {author} {\bibfnamefont {M.}~\bibnamefont {Shearer}}, \ and\
  \bibinfo {author} {\bibfnamefont {K.}~\bibnamefont {Zumbrun}},\ }\href@noop
  {} {\bibfield  {journal} {\bibinfo  {journal} {Euro. Jnl of Applied
  Mathematics}\ }\textbf {\bibinfo {volume} {12}},\ \bibinfo {pages} {253}
  (\bibinfo {year} {2001})}\BibitemShut {NoStop}%
\bibitem [{\citenamefont {{L}e{F}loch}\ and\ \citenamefont
  {Mohammadian}(2008)}]{lefloch2008}%
  \BibitemOpen
  \bibfield  {author} {\bibinfo {author} {\bibfnamefont {P.~G.}\ \bibnamefont
  {{L}e{F}loch}}\ and\ \bibinfo {author} {\bibfnamefont {M.}~\bibnamefont
  {Mohammadian}},\ }\href@noop {} {\bibfield  {journal} {\bibinfo  {journal}
  {J. Comp. Phys.}\ }\textbf {\bibinfo {volume} {227}},\ \bibinfo {pages}
  {4162} (\bibinfo {year} {2008})}\BibitemShut {NoStop}%
\bibitem [{\citenamefont {Bender}\ and\ \citenamefont
  {Orszag}(1999)}]{BenderOrszag}%
  \BibitemOpen
  \bibfield  {author} {\bibinfo {author} {\bibfnamefont {C.}~\bibnamefont
  {Bender}}\ and\ \bibinfo {author} {\bibfnamefont {S.}~\bibnamefont
  {Orszag}},\ }\href@noop {} {\emph {\bibinfo {title} {Advanced Mathematical
  Methods for Scientists and Engineers: Asymptotic Methods and Perturbation
  Theory}}}\ (\bibinfo  {publisher} {Springer},\ \bibinfo {year}
  {1999})\BibitemShut {NoStop}%
\bibitem [{\citenamefont {Liu}\ and\ \citenamefont
  {Zumbrun}(1995)}]{liuzumbrun95}%
  \BibitemOpen
  \bibfield  {author} {\bibinfo {author} {\bibfnamefont {T.-P.}\ \bibnamefont
  {Liu}}\ and\ \bibinfo {author} {\bibfnamefont {K.}~\bibnamefont {Zumbrun}},\
  }\href@noop {} {\bibfield  {journal} {\bibinfo  {journal} {Commun. Math.
  Phys.}\ }\textbf {\bibinfo {volume} {174}},\ \bibinfo {pages} {319} (\bibinfo
  {year} {1995})}\BibitemShut {NoStop}%
\bibitem [{\citenamefont {Sattinger}(1976)}]{sattinger76}%
  \BibitemOpen
  \bibfield  {author} {\bibinfo {author} {\bibfnamefont {D.~H.}\ \bibnamefont
  {Sattinger}},\ }\href@noop {} {\bibfield  {journal} {\bibinfo  {journal}
  {Advances in Math.}\ }\textbf {\bibinfo {volume} {22}},\ \bibinfo {pages}
  {312} (\bibinfo {year} {1976})}\BibitemShut {NoStop}%
\bibitem [{\citenamefont {Zumbrun}\ and\ \citenamefont
  {Serre}(1999)}]{zumbrun1999}%
  \BibitemOpen
  \bibfield  {author} {\bibinfo {author} {\bibfnamefont {K.}~\bibnamefont
  {Zumbrun}}\ and\ \bibinfo {author} {\bibfnamefont {D.}~\bibnamefont
  {Serre}},\ }\href@noop {} {\bibfield  {journal} {\bibinfo  {journal} {Indiana
  Math J.}\ }\textbf {\bibinfo {volume} {48}},\ \bibinfo {pages} {937}
  (\bibinfo {year} {1999})}\BibitemShut {NoStop}%
\bibitem [{\citenamefont {Giannuzzi}\ and\ \citenamefont
  {Stevie}(2005)}]{giannuzzi}%
  \BibitemOpen
  \bibinfo {editor} {\bibfnamefont {L.}~\bibnamefont {Giannuzzi}}\ and\
  \bibinfo {editor} {\bibfnamefont {F.}~\bibnamefont {Stevie}},\ eds.,\
  \enquote {\bibinfo {title} {Introduction to focused ion beams:
  Instrumentation, theory, techniques and practice},}\ \ (\bibinfo  {publisher}
  {Springer},\ \bibinfo {year} {2005})\ Chap.\ \bibinfo {chapter} {Appendix A:
  Ga Ion Sputter Yields}, pp.\ \bibinfo {pages} {329--331}\BibitemShut
  {NoStop}%
\bibitem [{\citenamefont {Yamamura}\ \emph {et~al.}(1983)\citenamefont
  {Yamamura}, \citenamefont {Itikawa},\ and\ \citenamefont
  {Itoh}}]{yamamura1983}%
  \BibitemOpen
  \bibfield  {author} {\bibinfo {author} {\bibfnamefont {Y.}~\bibnamefont
  {Yamamura}}, \bibinfo {author} {\bibfnamefont {Y.}~\bibnamefont {Itikawa}}, \
  and\ \bibinfo {author} {\bibfnamefont {N.}~\bibnamefont {Itoh}},\ }\href@noop
  {} {\enquote {\bibinfo {title} {Angular dependence of sputtering yields of
  monatomic solids},}\ }\bibinfo {howpublished} {Report No. IPPJ-AM-26}
  (\bibinfo {year} {1983})\BibitemShut {NoStop}%
\bibitem [{\citenamefont {Yamamura}\ \emph {et~al.}(1987)\citenamefont
  {Yamamura}, \citenamefont {M{\"o}ssner},\ and\ \citenamefont
  {Oechsner}}]{yamamura1987}%
  \BibitemOpen
  \bibfield  {author} {\bibinfo {author} {\bibfnamefont {Y.}~\bibnamefont
  {Yamamura}}, \bibinfo {author} {\bibfnamefont {C.}~\bibnamefont
  {M{\"o}ssner}}, \ and\ \bibinfo {author} {\bibfnamefont {H.}~\bibnamefont
  {Oechsner}},\ }\href@noop {} {\bibfield  {journal} {\bibinfo  {journal}
  {Radiat. Eff.}\ }\textbf {\bibinfo {volume} {103}},\ \bibinfo {pages} {25}
  (\bibinfo {year} {1987})}\BibitemShut {NoStop}%
\bibitem [{Note1()}]{Note1}%
  \BibitemOpen
  \bibinfo {note} {Technically this operator as defined is not analytic at 0,
  but it can be made analytic by an appropriate change of variables; see
  \protect \citet {bertozzi2001}.}\BibitemShut {Stop}%
\bibitem [{\citenamefont {Shraiman}(1986)}]{shraiman86}%
  \BibitemOpen
  \bibfield  {author} {\bibinfo {author} {\bibfnamefont {B.}~\bibnamefont
  {Shraiman}},\ }\href@noop {} {\bibfield  {journal} {\bibinfo  {journal}
  {Phys. Rev. Lett.}\ }\textbf {\bibinfo {volume} {57}},\ \bibinfo {pages}
  {325} (\bibinfo {year} {1986})}\BibitemShut {NoStop}%
\bibitem [{\citenamefont {Bernoff}(1988)}]{bernoff88}%
  \BibitemOpen
  \bibfield  {author} {\bibinfo {author} {\bibfnamefont {A.~J.}\ \bibnamefont
  {Bernoff}},\ }\href@noop {} {\bibfield  {journal} {\bibinfo  {journal}
  {Physica D}\ }\textbf {\bibinfo {volume} {30}},\ \bibinfo {pages} {363}
  (\bibinfo {year} {1988})}\BibitemShut {NoStop}%
\bibitem [{Note2()}]{Note2}%
  \BibitemOpen
  \bibinfo {note} {Eqn \protect \textup {\hbox {\mathsurround \z@ \protect
  \normalfont (\ignorespaces \ref {eq:O2}\unskip \@@italiccorr )}} is the
  expansion using a linearized 4th-order term $\propto \Delta ^2 h$; the
  expansion for the full nonlinear smoothing is similar but has more
  terms.}\BibitemShut {Stop}%
\end{thebibliography}%
